\relax
\documentstyle [art12]{article}
\begin {document}
\bibliographystyle {plain}

\title{\bf Mapping of the Calogero-Sutherland Model to the Gaussian Model}
\author {A. M. Tsvelik}
\maketitle
\begin {verse}
$Department~ of~ Physics,~ University~ of~ Oxford,~ 1~ Keble~ Road,$
\\
$Oxford,~OX1~ 3NP,~ UK$\\
\end{verse}
\begin{abstract}
\par
 The density-density correlation function of the Calogero-Sutherland model is represented as a Fourier transformation of the correlation function of 
bosonic exponents in the Gaussian model on a curved manifold.
\end{abstract}

PACS numbers: 05.30.-d, 71.10.+x
\sloppy
\par

 The Calogero-Sutherland model (CSM) describes a system of N nonrelativistic 
spinless particles interacting with inverse-square potential. The Hamiltonian has the 
following form
\begin{equation}
H = - \sum_{j = 1}^N\frac{\partial^2}{\partial x_j^2} + \sum_{i < j}\frac{2\lambda(\lambda - 1)}{d^2(x_i - x_j)}
\end{equation}
where $d(x) = |(L/\pi)\sin[\pi x/L]|$ and $L$ is the systems size. 

 At the special values $\lambda = 1/2, 1$ and 2, the CSM describes statistics of eigenvalues of random matrices belonging to one of the three Dyson's ensembles. Recently these special cases were intensely studied in connection with 
 the theory of universal spectral correlations in random systems \cite{borya}. 

 The spectrum of CSM was obtained by the so-called Asymptotic Bethe Ansatz \cite{sutherland} and turns out to be very simple due to the fact that 
CSM has  a trivial two-particle  scattering 
phase equal to $\phi(k_1,k_2) = \pi(\lambda - 1)\mbox{sign}(k_1 - k_2)$. 
As a consequence the CSM  can be thought of as an ideal gas of excitations obeying the fractional exclusion principle formulated in Ref.\cite{duncan}. 

 It is not surprising  that a theory with such a simple S-matrix 
has  particularly simple correlation functions. The exact  expression for the density-density correlation function  originally  conjectured by Haldane \cite{fdm} has been rigorously derived later using the method 
of Jack polynomials \cite{ha}.  It was established   
that for rational $\lambda = p/q$ this function has the following form:
\begin{eqnarray}
\langle 0|\rho(x,\tau)\rho(0,0)|0\rangle = C \prod_{i = 1}^q\left
(\int_0^{\infty}\mbox{d}x_i\right) \prod_{j = 1}^p\left
(\int_{-\lambda}^{0}\mbox{d}y_i\right)Q^2\cos Qx \mbox{e}^{ - E\tau}F(p,q,\lambda|\{x_i,y_j\})\label{func}\\
Q = 2\pi\rho\left(\sum_{i = 1}^qx_i - \lambda^{-1}\sum_{j = 1}^py_j\right), \nonumber\\
E = (2\pi\rho)^2\left[\sum_{i = 1}^q\epsilon(x_i) - \lambda^{-1}\sum_{j = 1}^p\epsilon(y_j)\right]
\end{eqnarray}
where $\rho = N/L$ is the average density of particles, $\epsilon(z) = z(z + \lambda)$ is the particles dispersion law, 
$C$ is the normalization factor whose eplicit form can be found in Ref.\cite{ha}   
and the function $F$ is given by 
\begin{eqnarray}
F(p,q,\lambda|\{x_i,y_j\}) = \prod_{i = 1}^q[\epsilon(x_i)]^{\lambda - 1}\prod_{a = 1}^p[\epsilon(y_i)]^{\lambda - 1}\frac{\left[\prod_{i < j}(x_i - x_j)^2\right]^{\lambda}\left[\prod_{a < b}(y_a - y_b)^2\right]^{1/\lambda}}{\prod_{i,a}(x_i - y_a)^{2}} \label{form}
\end{eqnarray}

 The main result of this Letter is the observation that the function $F(p,q,\lambda|\{x_i,y_j\})$ in  
Eq. (\ref{form}) can be represented  as the multipoint correlation function of 
the bosonic exponents of the chiral Gaussian model calculated on some curved manifold. The latter model has 
the following action:
\begin{eqnarray}
S = \frac{1}{8\pi}\int_{A}\mbox{d}\tau\mbox{d}x\partial_x\phi[\mbox{i}\partial_{\tau}\phi + \partial_x\phi] \label{gauss}
\end{eqnarray}
where $A$ is an area of two dimensional plane where the field $\phi$ 
is defined. 

 Thus the entire correlation function in real space is represented as a sort of Fourier transformation of the correlation function of the free field theory 
defined in {\it momentum space}. 
Such representations are  known in the theory of integrable models where 
they are  called ``momentum space bosonization'' \cite{leclair}. 
A similar approach to the CSM was taken by Khveshchenko\cite{dima}. 
He, however,  
expressed the CSM form-factors in terms of correlation functions of an interacting bosonic model. 

A remarkable property of the action (\ref{gauss}) is that the correlation functions of bosonic exponents are holomorphic, i. e. they 
depend only on the variable $z = \tau + \mbox{i}x$. In particular, the 
two-point correlation function {\it on the infinite plane} is equal to 
\begin{eqnarray}
\langle \mbox{e}^{\mbox{i}\beta\phi(z_1)}\mbox{e}^{- \mbox{i}\beta\phi(z_2)}\rangle = (z_{12})^{- 2\beta^2}
\end{eqnarray}
A multi-point correlation function {\it on the infinite plane} is given by 
\begin{eqnarray}
F(z_1\beta_1, ...z_N,\beta_N) \equiv
\langle \mbox{e}^{\mbox{i}\beta_1\phi(1)}...\mbox{e}^{\mbox{i}\beta_N\phi(N)}\rangle = \left[\prod_{i < j}(z_{ij})^{2\beta_i\beta_j}\right]\delta_{\sum\beta_i,0} \label{multi}
\end{eqnarray}

 Suppose now that transformation $z(w)$ maps the area $A$ onto infinite plane. Then correlation functions on the area $A$ can be obtained from the correlation function on infinite plane by the following transformation:
\begin{eqnarray}
F(w_1\beta_1, ...w_N,\beta_N)_A = F[z(w_1)\beta_1, ...z(w_N),\beta_N]\prod_i(\frac{\mbox{d}z}{\mbox{d}w_i})^{\beta_i^2} \label{trans}
\end{eqnarray}

 Comparing Eqs.(\ref{multi}) and (\ref{trans}) with Eqs.(\ref{func}) and (\ref{form}) we conclude that Eq.(\ref{func}) can be written in the following form:
\begin{eqnarray}
\langle 0|\rho(x,\tau)\rho(0,0)|0\rangle = \nonumber\\
C \prod_{i = 1}^q\left
(\int_{C_1}\mbox{d}w_i^+\right) \prod_{j = 1}^p\left
(\int_{C_2}\mbox{d}w_j^-\right)Q^2\cos Qx \mbox{e}^{ - E\tau}\langle \prod_{i = 1}^q \mbox{e}^{\mbox{i}\sqrt\lambda\phi(w^+_i)}\prod_{j = 1}^p\mbox{e}^{- \mbox{i}\phi(w^-_j)/\sqrt\lambda}\rangle_A \label{res}
\end{eqnarray}
where the transformation which maps area $A$ onto infinite plane is determined by the equation
\begin{equation}
\frac{\mbox{d}z}{\mbox{d}w} = \epsilon(z) \label{trans}
\end{equation}
that is 
\begin{equation}
z = \frac{\lambda}{\mbox{e}^{ - \lambda w} - 1} \label{conf}
\end{equation}
Here we remark that Eq.(\ref{trans}) for general $\epsilon(z)$ describes a general relationship between rapidities $w$ and momenta $z$ in integrable theories of ``magnets'' and therefore is likely to have a universal status. In our 
case the transformation (\ref{conf}) maps the contours of 
integration $C_1, C_2$ on the original contours. It follows from (\ref{conf}) 
that $C_1$ coincides with the negative part of the real axis and $C_2$ runs from minus to plus infinity parallel to the real axis: $\Im m w^- = \pi/\lambda$. 

 Notice that the requirement $\lambda = p/q$ follows naturally from the 
``electroneutrality'' condition $\sum_i\beta_i = 0$ (see Eq.(\ref{multi})). 

 Thus we have established a one-to-one correspondence between the 
correlation function of the CSM and the correlation function of the theory 
of free bosonic field. We can use this equivalence to 
construct 
generalized CS models. As an  example we shall consider 
the model in which momentum space bosonization is described  by the theory with the action 
\begin{eqnarray}
S = \frac{1}{8\pi}\int_{A}\mbox{d}\tau\mbox{d}x\sum_{a = 1}^{M - 1}\partial_x\phi_a[\mbox{i}\partial_{\tau}\phi_a + \partial_x\phi_a]
\end{eqnarray}
with bosonic exponents defined as
\begin{equation}
O_{+,i}(z) = \mbox{e}^{\mbox{i}\lambda^{1/2}{\vec\beta_i\vec\phi}(z)}, \: O_{-,j} = \mbox{e}^{\mbox{i}\lambda^{-1/2}{\vec\gamma_j\vec\phi}(z)} 
\end{equation}
where $({\vec\beta_i\vec\gamma_j}) = \delta_{ij}$ and ${\vec\beta_i}$ are the simple root vectors of $A_{M - 1}$ defined by $(\vec\beta_i\vec\beta_j) = 2\delta_{ij} - (\delta_{i,j-1} + \delta_{i,j+1}))$.  Since individual root 
vectors are not parallel and only the  sums of 
all vectors ${\vec\beta}$ and ${\vec\gamma}$ are, 
the electroneutrality condition for the multi-point correlation function of particle $O_{+}$- and hole operators $O_{-}$ is fulfilled only if the correlation function contains  products of all colours. That is we have $\lambda = (M - 1)p/2q$ (extra 2 factor comes from the different definition of $\vec\beta_i$ for $M = 2$ and $M > 2$). The correlation function is equal to 
\begin{eqnarray}
\langle \prod_{i = 1}^{q(M - 1)}O_{+,i}(z^+_i)\prod_{a = 1}^{p(M - 1)}
O_{-,j}(z^-_j)\rangle = \nonumber\\
\frac{\prod_{i < j}\left[(z_i^+ - z_j^+)^2\right]^{\lambda{\vec\beta_i\vec\beta_j}}\prod_{a < b}\left[(z_a^- - z_b^-)^2\right]^{\lambda^{-1}{\vec\gamma_a\vec\gamma_b}}}{\prod_{a,i}(z^+_i - z^-_j)^2} 
\end{eqnarray}
It is not equal to zero provided 
\begin{eqnarray}
\lambda q\sum_{i = 1}^{M - 1}\vec\beta_i = p\sum_{j = 1}^{M - 1}\vec\gamma_j
\end{eqnarray}
or equivalently  $\lambda = (M - 1)p/2q$. To get the correlation function in real space one should perform the conformal transformation (\ref{conf}) and substitute the result into Eq.(\ref{res}). 

Another  set of generalized CS models can be obtained from  the  chiral 
Gaussian models introduced by Haldane to describe edge state in 
Fractional Quantum Hall effect\cite{haldane}:
\begin{eqnarray}
S = \frac{1}{8\pi}\int_{A}\mbox{d}\tau\mbox{d}x\partial_x\phi_a[K^{-1}_{ab}\mbox{i}\partial_{\tau}\phi_b + \partial_x\phi_a]
\end{eqnarray}
where the matrix $K$ is described in the original publication. 

We are grateful to B. L. Altshuler and B. D. Simons  for the inspirational  
discussions and
interest in the work. This research was supported in part by the National Science Foundation under Grant No. PHY94-07194.

\end{document}